\documentclass[%
  reprint,
superscriptaddress,
showpacs,%preprintnumbers,
bibnotes,
 amsmath,amssymb,
]{revtex4-1}

\usepackage[breaklinks=true,colorlinks=true,linkcolor=blue,urlcolor=blue,citecolor=blue]{hyperref}

\usepackage{floatrow}
\usepackage{braket}
\usepackage{graphicx}% Include figure files
\usepackage{dcolumn}% Align table columns on decimal point
\usepackage{bm}% bold math
\begin{document}

\preprint{APS/123-QED}

\title{Permutation blocking path integral Monte Carlo approach to the uniform electron gas at finite temperature}% Force line breaks with \\

\author{Tobias Dornheim}
 \email{dornheim@theo-physik.uni-kiel.de}
%  \altaffiliation[Also at ]{Physics Department, XYZ University.}%Lines break automatically or can be forced with \\
\author{Tim Schoof}%
\author{Simon Groth}

\affiliation{%
 Institut f\"ur Theoretische Physik und Astrophysik, Christian-Albrechts-Universit\"at, Leibnizstrasse 15, Kiel D-24098, Germany
}%

\author{Alexey Filinov}
\affiliation{%
 Institut f\"ur Theoretische Physik und Astrophysik, Christian-Albrechts-Universit\"at, Leibnizstrasse 15, Kiel D-24098, Germany
}%
\affiliation{%
 Joint Institute for High Temperatures RAS, Izhorskaya Str. 13, 125412 Moscow,
Russia
}%

\author{Michael Bonitz}

\affiliation{%
 Institut f\"ur Theoretische Physik und Astrophysik, Christian-Albrechts-Universit\"at, Leibnizstrasse 15, Kiel D-24098, Germany
}%

% \collaboration{MUSO Collaboration}%\noaffiliation
%  \homepage{http://www.Second.institution.edu/~Charlie.Author}
% \affiliation{
%  Second institution and/or address\\
%  This line break forced% with \\
% }%
% \affiliation{
%  Third institution, the second for Charlie Author
% \affiliation{%
%  Authors' institution and/or address\\
%  This line break forced with \textbackslash\textbackslash
% }%
% 
% \collaboration{CLEO Collaboration}%\noaffiliation

\date{\today}% It is always \today, today,

\begin{abstract}
The uniform electron gas (UEG) at finite temperature is of high current interest due to its key relevance for many applications
including dense plasmas and laser excited solids. In particular, density functional theory heavily relies on accurate thermodynamic data for the UEG. Until recently, the only existing first-principle results had been obtained for $N=33$ electrons with restricted path integral Monte Carlo (RPIMC), for low to moderate density,
%which is limitted to moderate density, 
 $r_s = \overline{r}/a_\textnormal{B} \gtrsim 1$. This data has been complemented by Configuration path integral Monte Carlo (CPIMC) simulations for $r_s \leq 1$ that substantially deviate from
%however, called into question the validity of 
RPIMC towards smaller $r_s$ and low temperature.
In this work, we 
%partly resolve this discrepancy 
present results from an independent third method---the recently developed permutation blocking path integral Monte Carlo (PB-PIMC) approach [T. Dornheim \textit{et al.}, NJP \textbf{17}, 073017 (2015)] which we extend to the UEG.
Interestingly, PB-PIMC allows us to perform simulations over the entire density range down to half the Fermi temperature ($\theta=k_BT/E_F=0.5$)
and, therefore, to compare our results to both aforementioned methods.
While we find excellent agreement with CPIMC, where results are available, we observe deviations from RPIMC that are beyond the statistical errors and increase with density.
% \begin{description}
% Secondary publications and information retrieval purposes.
% \item[PACS numbers]
% May be entered using the \verb+\pacs{#1}+ command.
% You may use the \texttt{description} environment to structure your abstract;
% use the optional argument of the \verb+\item+ command to give the category of each item. 
% \end{description}
\end{abstract}

\maketitle

%\tableofcontents

\section{\label{intro}Introduction}
Over the last years, there has been an increasing interest in the thermodynamic properties of degenerate electrons in the quantum mechanical regime.
Such information is vital for the description of highly compressed matter \cite{fletcher,kraus,regan}, 
including plasmas in laser fusion experiments \cite{lindl,hu,hurricane,nora,gomez,schmit}
and in compact stars and planet cores \cite{knudson,militzer,nettelmann}.
In addition, the widespread density functional theory (DFT) approach crucially depends on the availability
of accurate quantum Monte Carlo (QMC) data for the exchange correlation energy of the UEG, hitherto at zero temperature \cite{alder,fci}.
However, in recent years more and more applications with highly excited electrons have emerged, which require to go beyond ground state DFT.
Hence, there exists a high current need for an ab-initio thermodynamic description of the UEG at finite $T$.

The widely used path integral Monte Carlo (PIMC) method, e.g.\ \cite{cep}, is a powerful tool for the ab-initio simulation of both distinguishable
particles (often referred to as ``boltzmannons'', e.g.\ \cite{mil,clark}) and bosons and allows for quasi exact
results for up to $N\sim10^3$ particles at finite temperature \cite{bon,bon2}.
However, the application of PIMC to fermions is hampered by the notorious fermion sign problem (FSP), e.g.\ \cite{loh}, which might render
even small systems unfeasible for state of the art QMC methods and is known to be NP-hard for a given representation \cite{troyer}.
With increasing degeneracy effects, permutation cycles with opposite signs nearly cancel each other and the statistical uncertainty 
grows exponentially. Hence, standard PIMC cannot provide the desired results without further improvement.
Brown \textit{et al.} \cite{brown} have presented the first finite temperature results for the UEG down to $r_s=1$ using
restricted PIMC (RPIMC) \cite{node},
a popular approach to extend PIMC to higher degeneracy, that is, lower temperature and higher density.
To avoid the FSP, this method requires explicit knowledge of the nodal surface of the density matrix, which is, in general, unknown
and one has to rely on approximations. The use of the ideal nodes for a nonideal system appears to be problematic,
as has been shown for the case of hydrogen \cite{hyd1,hyd2}. In addition, it has been shown analytically that RPIMC does not reproduce
the exact limit of the ideal Fermi gas ($r_s\to0$) \cite{vfil1,vfil2}. Therefore, the quality of the RPIMC data remains unclear.
Indeed, recent configuration PIMC (CPIMC) \cite{tim1,tim2} results for the highly degenerate UEG
by Schoof \textit{et al.} \cite{prl} have revealed a significant disagreement between the two methods at small $r_s$ and low temperature.
While the first application of a novel density matrix QMC (DMQMC) approach \cite{blunt} to the UEG for four particles reports
 excellent agreement with CPIMC \cite{malone},
additional simulations of larger systems are needed to resolve the discrepancy towards RPIMC.
For completeness, we mention that QMC results by Filinov \textit{et al.} \cite{vfil4} cannot be used as a benchmark 
due to the different treatment of the homogeneous positive background and a different account of the long-range Coulomb interaction \cite{yakub1,yakub2} than the usual Ewald summation. In this situation an independent third first-principle method, capable to treat WDM parameters, would be highly desirable.

In this work we, therefore, investigate the applicability of the recently developed permutation blocking PIMC (PB-PIMC) approach \cite{dornheim} to the uniform electron gas.
The basic idea behind PB-PIMC is to combine antisymmetric imaginary time propagators  \cite{det1,det2,det3}, i.e., determinants, between all ``time slices'' with a higher order factorization of the density matrix \cite{ho1,ho2,ho3,ho4}.
This means that each particle is represented by a ``path'' consisting of $3 \times P$ coordinates (``beads''), where $P$ is the number of high-temperature factors (or propagators).
The application of determinants leads to a relieve of the FSP by an effective cancellation of positive and negative terms in the partition function,
which belong to permutation cycles of different parity in standard PIMC. However, since the blocking is most effective
if the thermal wavelength of a single propagator is of the same order as the mean interparticle distance, 
it is crucial to employ a higher order factorization scheme which allows for sufficient accuracy with only a few time slices.

\begin{figure}
      \centering
      \includegraphics[width=0.98\textwidth]{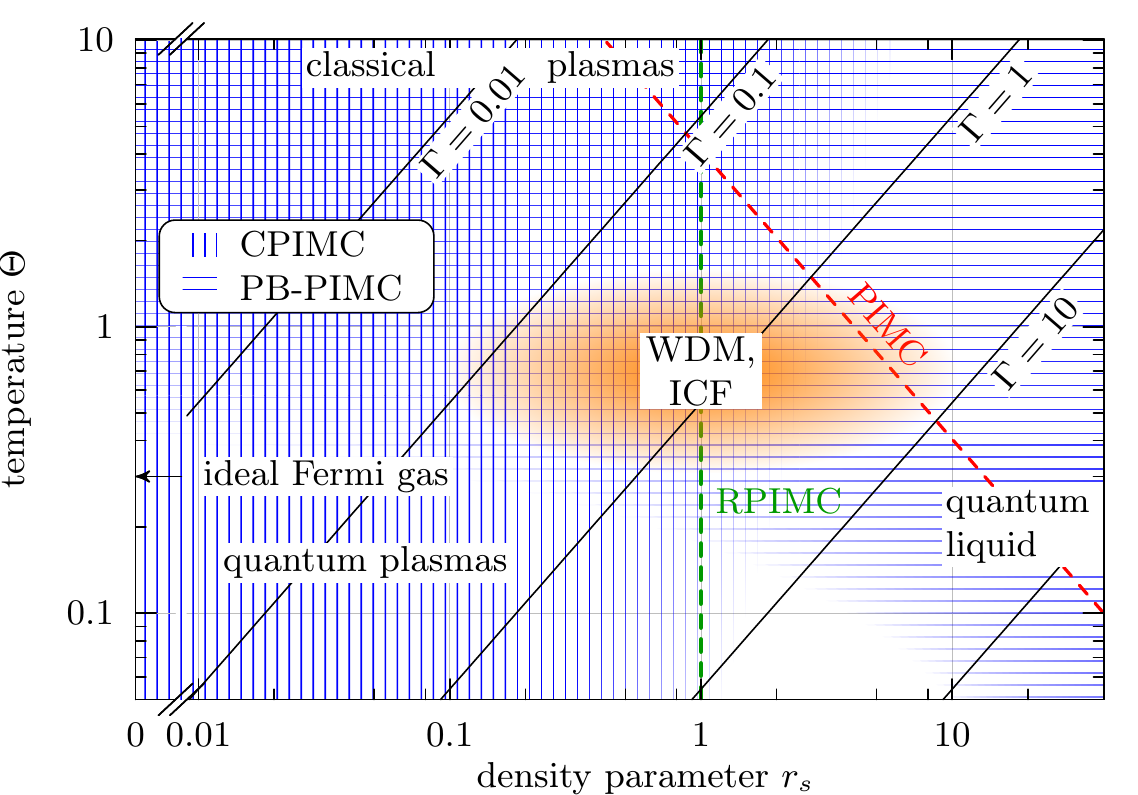}
      \caption{Density-temperature plain around the warm dense matter (WDM) regime. PB-PIMC significantly extends the range of applicability of standard PIMC (qualitatively shown by the red dashe line)
      towards lower temperature and higher density while CPIMC is applicable to the highly degenerate and weakly nonideal UEG \cite{prl}.
      RPIMC data \cite{brown} are available for $r_s \ge 1$.
%      , although the quality of the results is questionable towards high degeneracy. 
      The orange area marks the conditions of WDM and inertial confinement fusion (ICF) \cite{hu}.
      }
      \label{sketch}
  \end{figure}

The details of our PB-PIMC scheme are described in section \ref{method}, after a brief introduction of the employed model \ref{model}.
In section \ref{sconvergence}, we present our simulation results starting with a detailed investigation of the convergence behavior with respect
to the factorization of the density matrix.
We proceed  by simulating $N=33$ spin-polarized electrons, which is a commonly used model system of the UEG, see section \ref{sdensity}.
Interestingly, our PB-PIMC approach allows us to obtain accurate results over the entire density range and, therefore, to make a comparison
with the pre-existing RPIMC and CPIMC results for the UEG.
Finally, in section \ref{td} we investigate the applicability of our method with respect to the temperature.
We find that PB-PIMC, in combination with CPIMC, allows for the simulation of the UEG over a broad parameter range, which includes
the physically most interesting regime of warm dense matter, cf.\ Fig.\ \ref{sketch}.

\section{\label{theory}Theory}
\subsection{\label{model}Model Hamiltonian}
The uniform electron gas, often referred to as ``Jellium'', is a model description of Coulomb interacting electrons with a neutralizing background
of positive charges which are uncorrelated and homogeneously distributed.
To describe an infinite system based on a finite number of particles, one implements periodic boundary conditions and
includes the interaction of the $N$ electrons in the main cell with all their images via Ewald summation.
Following the notation from \cite{fraser}, we express the Hamiltonian of the $N$ electron UEG (in atomic units) as
 \begin{eqnarray}
  \nonumber \hat{H} = -\frac{1}{2} \sum_{i=1}^N \nabla^2_i + \frac{1}{2} \sum_{i=1}^N\sum_{j\ne i}^N e^2 \Psi( \mathbf{r}_i, \mathbf{r}_j) + \frac{N e^2}{2}\xi \quad ,
 \end{eqnarray}
with $\xi$ being the Madelung constant and the periodic Ewald pair potential
\begin{eqnarray}
\nonumber \Psi(\mathbf{r}, \mathbf{s} ) &=& \frac{1}{V} \sum_{ \mathbf{G} \ne 0 } \frac{ e^{-\pi^2\mathbf{G}^2/\kappa^2} e^{2\pi i \mathbf{G}(\mathbf{r}-\mathbf{s})} }{ \pi\mathbf{G}^2}
 \\ \label{pair} &-& \frac{\pi}{\kappa^2 V} + \sum_\mathbf{R} \frac{ \textnormal{erfc}( \kappa | \mathbf{r}-\mathbf{s} + \mathbf{R} | ) }{ |\mathbf{r}-\mathbf{s}+\mathbf{R} | } \quad .
\end{eqnarray}
Here $\mathbf{R}=\mathbf{n}_1L$ and $\mathbf{G}=\mathbf{n}_2/L$ denote the real and reciprocal space lattice vectors, respectively, with the box length $L$ and volume $V=L^3$.
The specific choice of the Ewald parameter $\kappa$ does not influence the outcome of Eq.\ (\ref{pair}) and, therefore, can be used to optimize the 
convergence. PB-PIMC requires explicit knowledge of all forces in the system, and the force between the electrons $i$ and $j$ can be obtained from
 \begin{eqnarray}
  \mathbf{F}_{ij} = - \nabla_i \Psi( \mathbf{r}_i, \mathbf{r}_j ) \quad . \label{iforce}
 \end{eqnarray}
The evaluation of Eq.\ (\ref{iforce}) is relatively straightforward and we find
\begin{eqnarray}
 \mathbf{F}_{ij} &=&  \frac{2}{V} \sum_{\mathbf{G}\ne 0}\left( \frac{ \mathbf{G} }{ \mathbf{G}^2 }  \textnormal{sin}\left[ 2\pi\mathbf{G}(\mathbf{r}_i - \mathbf{r}_j)\right]e^{- {\pi^2\mathbf{G}^2}/{\kappa^2}} \right)  \nonumber \\ &+&
 \sum_{\mathbf{R}} \frac{ \mathbf{r}_i - \mathbf{r}_j + \mathbf{R} }{ \alpha^3} \left( \textnormal{erfc}(\kappa\alpha) + \frac{2\kappa\alpha}{\sqrt{\pi}}e^{-\kappa^2\alpha^2} \right)
\quad , \nonumber
 \end{eqnarray}
 with the definition $\alpha = | \mathbf{r}_i - \mathbf{r}_j + \mathbf{R} |$.
 
 \subsection{\label{method}Simulation method}
 To calculate canonical expectation values with the PB-PIMC approach \cite{dornheim}, we write the partition function in coordinate representation
 as 
 \begin{eqnarray}
\label{z} Z = \frac{1}{N!} \sum_{\sigma\in S_N} \textnormal{sgn}(\sigma) \int \textnormal{d}\mathbf{R}\ \bra{\mathbf{R}} e^{-\beta\hat{H}} \ket{\hat{\pi}_\sigma\mathbf{R}} \quad ,
\end{eqnarray}
 with $\mathbf{R}={\mathbf{r}_1,\dots,\mathbf{r}_N}$ containing the coordinates of all electrons, $\hat{\pi}_\sigma$ denoting the 
 exchange operator which corresponds to a specific element $\sigma$ from the permutation group $S_N$ and $\beta=1/k_\textnormal{B}T$.
 For the next step, we make use of the usual group property of the density matrix in Eq.\ (\ref{z}) and arrive at an expression for $Z$ which requires
 the evaluation of $P$ density matrices at $P$ times higher temperature. However, instead of the primitive approximation $e^{-\epsilon\hat{H}}\approx e^{-\epsilon\hat{K}}e^{-\epsilon\hat{V}}$, with
 $\epsilon = \beta / P$ being the imaginary time step of a single propagator and the kinetic and potential contributions to the Hamiltonian $\hat{K}$ and $\hat{V}$, respectively, we
 use the fourth order factorization \cite{ho2,ho3}
 \begin{eqnarray}
\nonumber e^{-\epsilon\hat{H}} \approx & & e^{-v_1\epsilon\hat{W}_{a_1}} e^{-t_1\epsilon\hat{K}}  e^{-v_2\epsilon\hat{W}_{1-2a_1}} \\  & & \times e^{-t_1\epsilon\hat{K}} e^{-v_1\epsilon\hat{W}_{a_1}} e^{-2t_0\epsilon\hat{K}} \quad . \label{fop}
\end{eqnarray}
The $\hat{W}$ operators in Eq.\ (\ref{fop}) denote a modified potential, which combines $\hat{V}$ with double commutator terms of the form
 \begin{eqnarray}
 [[\hat{V},\hat{K}],\hat{V}] = \frac{\hbar^2}{m} \sum_{i=1}^N |\mathbf{F}_i|^2 \quad ,
\end{eqnarray}
and, therefore, requires the evaluation of all forces on each particle, $\mathbf{F}_i = -\nabla_i V(\mathbf{R})$.
Our final result for the partition function is given by
\begin{eqnarray}
\label{finalz} Z &=& \frac{1}{(N!)^{3P}} \int \textnormal{d}\mathbf{X} \prod_{\alpha=0}^{P-1} e^{-\epsilon\tilde V_\alpha}e^{-\epsilon^3u_0\frac{\hbar^2}{m}\tilde F_\alpha} \\
 & & \textnormal{det}(\rho_\alpha)\textnormal{det}(\rho_{\alpha A})\textnormal{det}(\rho_{\alpha B}) \quad , \nonumber 
\end{eqnarray}
 with the definition of the potential and force terms
 \begin{eqnarray}
 \tilde V_\alpha &=& v_1 V(\mathbf{R}_\alpha) + v_2 V(\mathbf{R}_{\alpha A}) + v_1 V(\mathbf{R}_{\alpha B}) \quad ,\\
 \nonumber \tilde F_\alpha &=& \sum_{i=1}^N \left( a_1 |\mathbf{F}_{\alpha,i}|^2 + (1-2a_1) |\mathbf{F}_{\alpha A,i}|^2 + a_1 |\mathbf{F}_{\alpha B,i}|^2 \right)  \quad ,
\end{eqnarray}
 and the diffusion matrices
 \begin{eqnarray}
 \nonumber \rho_\alpha(i,j) = \lambda_{t_1\epsilon}^{-D} \sum_\mathbf{n} \textnormal{exp} \left( -\frac{\pi}{\lambda^2_{t_1\epsilon}} ( \mathbf{r}_{\alpha,j} - \mathbf{r}_{\alpha A,i} + \mathbf{n}L)^2 \right) \quad ,
\end{eqnarray}
with $D$ being the dimensionality, see e.g.\ \cite{det1}.
Eq.\ (\ref{finalz}) contains two free coefficients, $t_0$ and $a_1$, which can be used for optimization, cf.\ Fig.\ \ref{parameter},
and the integration is carried out over $3P$ sets of coordinates, $\textnormal{d}\mathbf{X} =\textnormal{d}\mathbf{R}_{0}\dots\textnormal{d}\mathbf{R}_{P-1}\textnormal{d}\mathbf{R}_{0A}\dots\textnormal{d}\mathbf{R}_{P-1A}\textnormal{d}\mathbf{R}_{0B}\dots\textnormal{d}\mathbf{R}_{P-1B}$.
Instead of explicitly sampling each permutation individually, as in standard PIMC, we combine configuration weights of both positive and negative sign  
in the determinants, which leads to a cancellation of terms and, therefore, an effective blocking of permutations. 
When the thermal wavelength of a single time slice, $\lambda_{t_1\epsilon} = \sqrt{2\pi\epsilon t_1\hbar^2/m}$, is comparable 
to the mean interparticle distance, the effect of the blocking is most pronounced and the average sign in our simulations is significantly increased.
However, with an increasing number of propagators $P$, $\lambda_{t_1\epsilon}$ decreases and, eventually, the blocking will have no effect and the sign
converges towards the sign from standard PIMC. Hence, it is crucial to employ the high order factorization from Eq.\ (\ref{fop}), which allows
for reasonable accuracy even for only two or three propagators.
We simulate the canonical probability distribution defined by Eq.\ (\ref{finalz}) using the Metropolis algorithm \cite{metropolis}
and refer to \cite{dornheim} for a more detailed description of the PB-PIMC method.

\subsection{Energy estimator}
The consideration of periodicity in the diffusion matrices requires minor modifications in the energy estimator presented in \cite{dornheim},
which can be derived from the partition function via the familiar relation
\begin{eqnarray}
 E = - \frac{1}{Z} \frac{\partial Z}{\partial \beta} \quad . \label{tde}
\end{eqnarray}
Inserting the expression from Eq.\ (\ref{finalz}) into (\ref{tde}) and performing a lengthy but straightforward calculation leads to
\begin{eqnarray}
 \nonumber E &=& \frac{1}{P}\sum_{k=0}^{P-1}\left( \tilde V_k + 3\epsilon^2u_0\frac{\hbar^2}{m}\tilde F_k \right) + \frac{3DN}{2\epsilon}
 \\ &-& \nonumber \sum_{k=0}^{P-1}\sum_{\kappa=1}^N\sum_{\xi=1}^N\left( \frac{ \pi\eta^{k}_{\kappa\xi} }{ \epsilon P \lambda_{t_1\epsilon}^2 } 
 + \frac{ \pi\eta^{kA}_{\kappa\xi} }{ \epsilon P \lambda_{t_1\epsilon}^2 } + \frac{ \pi \eta_{\kappa\xi}^{kB} }{ \epsilon P \lambda_{2t_0\epsilon}^2} \right)
 \quad ,
\end{eqnarray}
 with the definition
 \begin{eqnarray}
  \eta^k_{\kappa\xi} &=&  \frac{\left( \rho_k^{-1}\right)_{\kappa\xi} }{\lambda_{t_1\epsilon}^D } \nonumber
 \sum_\mathbf{n} \textnormal{exp}\left[ -\frac{\pi}{\lambda_{t_1\epsilon}^2}(\mathbf{r}_{k,\kappa}-\mathbf{r}_{kA,\xi}+L\mathbf{n})^2\right] \\
 & &  (\mathbf{r}_{k,\kappa}-\mathbf{r}_{kA,\xi}+L\mathbf{n})^2 \quad .
 \end{eqnarray}
 For completeness, we note that the total energy $E$ splits into the kinetic and potential contribution, $K$ and $V$, in precisely
 the same way as before \cite{dornheim}.

\section{Results\label{results}}
%  We begin the discussion of our simulation results by investigating the convergence of the energy with the number of imaginary time propagators $P$.

 \subsection{Convergence\label{sconvergence}}
\begin{figure}
      \centering
      \includegraphics[width=0.98\textwidth]{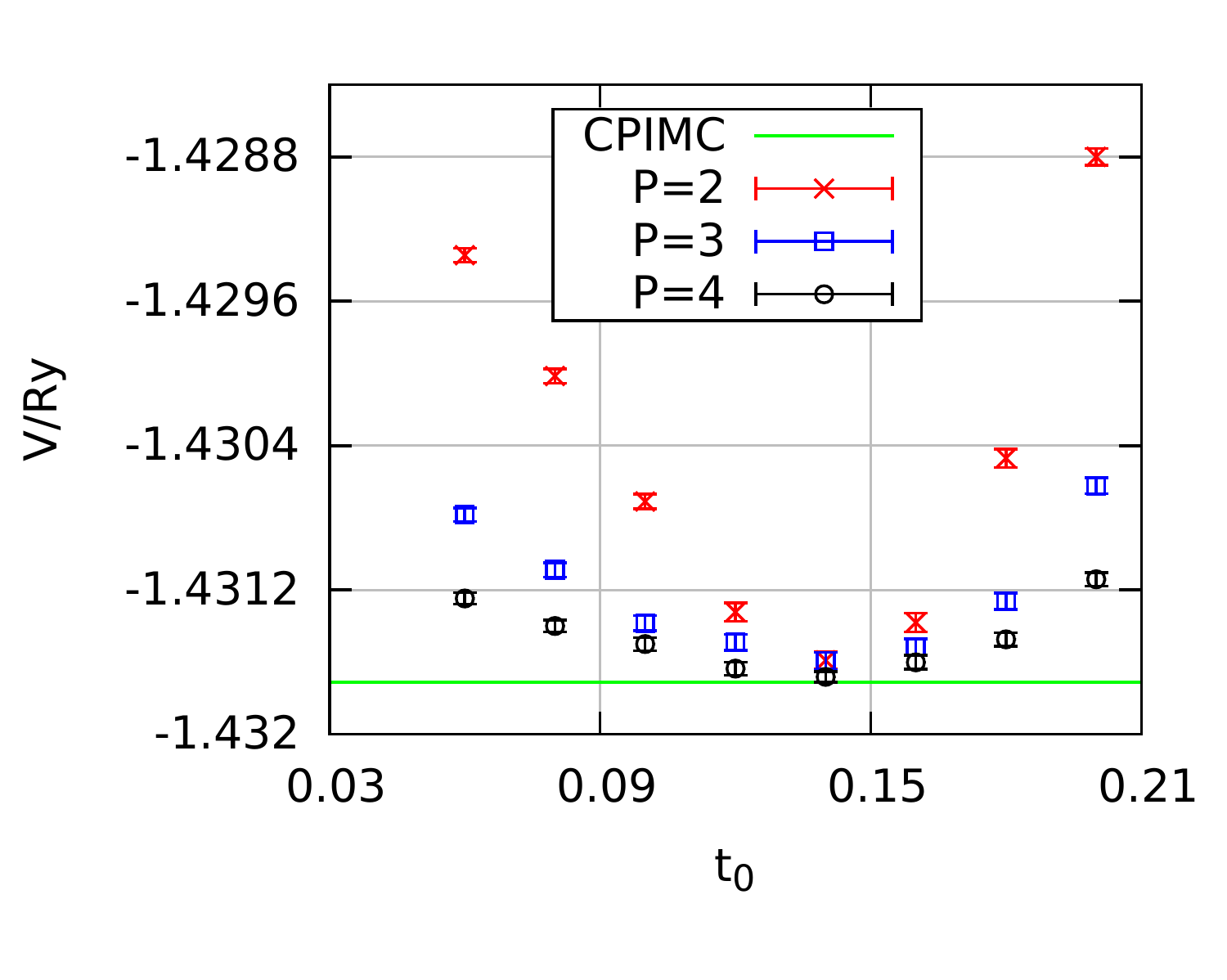}
      \caption{Influence of the relative interslice spacing $t_0$ for $N=4$, $r_s=4$ and $\theta=0.5$ on the convergence of the propagator. 
      The exact result known from CPIMC (green line) is compared to the PB-PIMC results for $P=2$, $P=3$ and $P=4$ for 
      the fixed free parameter $a_1=0.33$ over the entire $t_0$ range. The optimal value is located around $t_0=0.14$.
      }
      \label{parameter}\vspace*{-0.5cm}
  \end{figure}

    \begin{figure}
      \centering
      \includegraphics[width=0.98\textwidth]{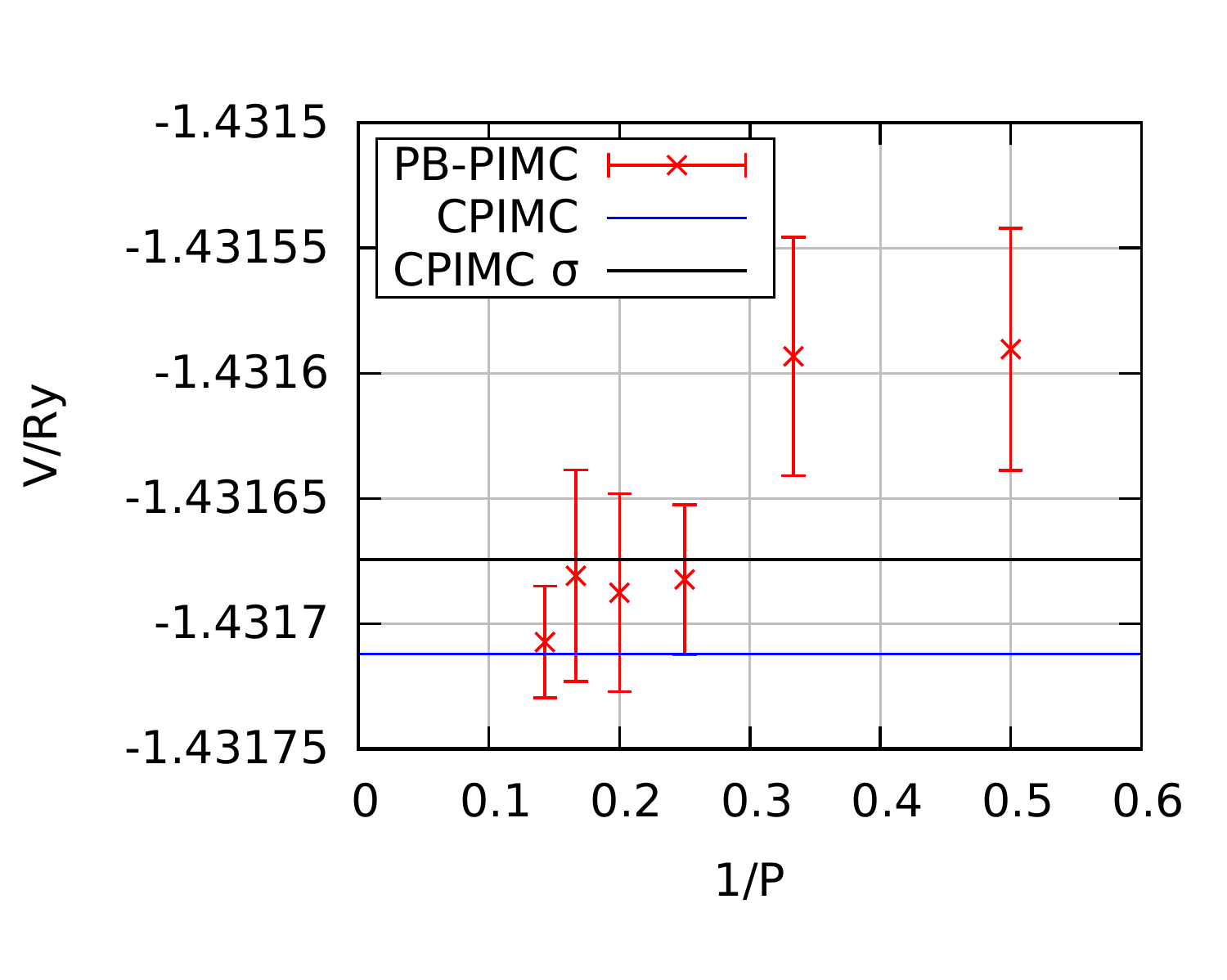}\\
         \vspace*{-0.8cm}
      \includegraphics[width=0.98\textwidth]{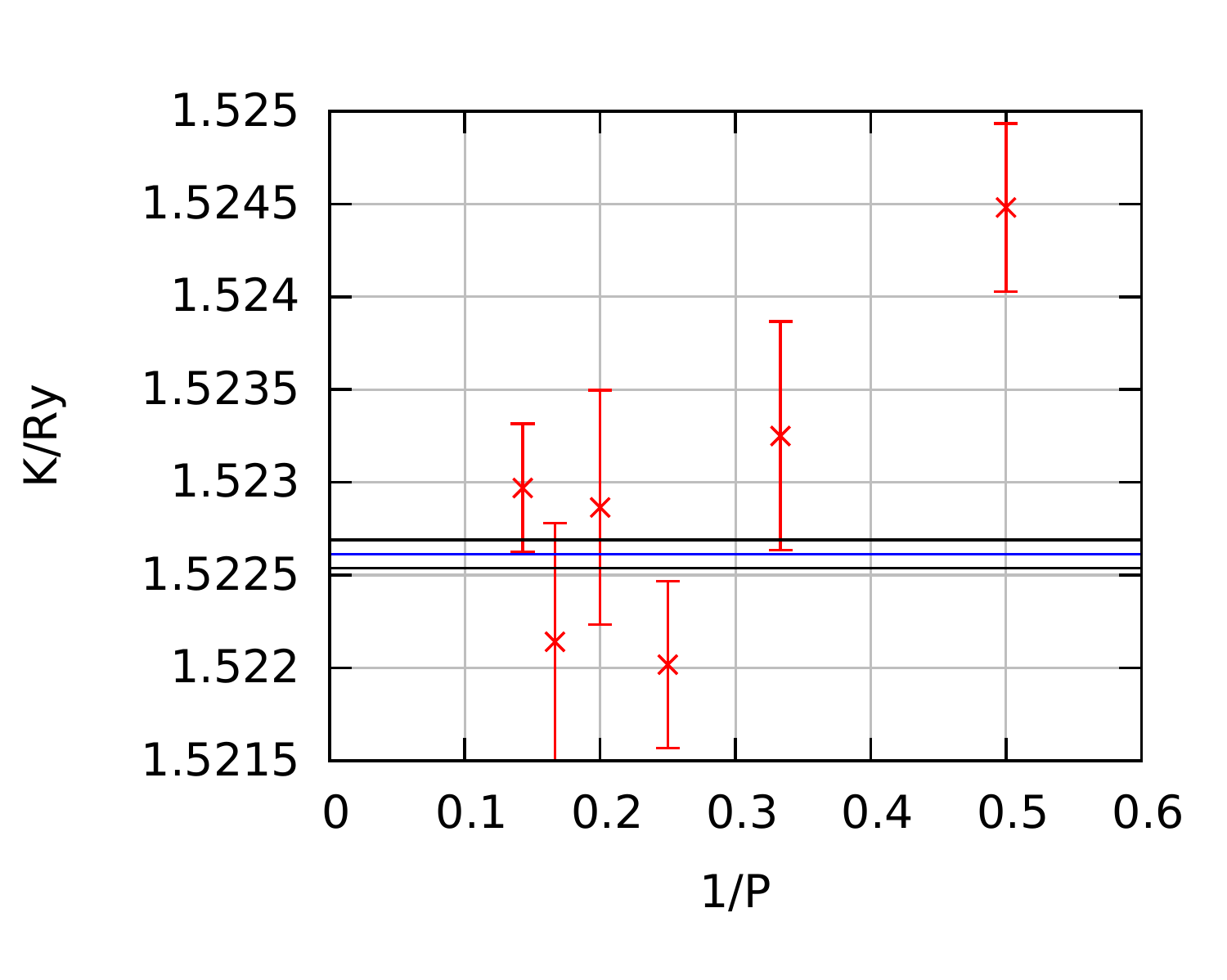}
      \caption{Convergence of the potential (top) and kinetic (bottom) energy for $N=4$, $r_s=4$ and $\theta=0.5$ with $t_0=0.14$ and $a_1=0.33$. 
      In the top panel, the potential energy $V$ is plotted versus the inverse number of propagators $P^{-1}\propto\epsilon$
      and the PB-PIMC results are compared to the exact value known from CPIMC. The bottom panel shows the same information for the kinetic energy $K$.
      }
      \label{convergence}\vspace*{-0.5cm}
  \end{figure}

  We begin the discussion of our simulation results by investigating the convergence of the energy with the number of imaginary time propagators $P$.
To enhance the performance, the free parameters from the propagator, $a_1$ and $t_0$, can be optimized.
  In Fig.\ \ref{parameter}, we choose $a_1=0.33$, which corresponds to equally weighted forces on all time slices, and plot the potential energy $V$, calculated with $P=2$, $P=3$ and $P=4$, versus $t_0$ over the entire possible range
  for a benchmark system of $N=4$ spin-polarized electrons with $\theta=0.5$ and $r_s=4$. To asses the accuracy, we compare these results with the exact energy known
  from CPIMC (green line). Evidently, the optimal choice for this free parameter is located around $t_0=0.14$, which is consistent with previous findings by Sakkos \textit{et al.} \cite{ho3}
  and the application of PB-PIMC to electrons in a quantum dot \cite{dornheim}. 
  For completeness, we mention that the kinetic energy $K$
  exhibits the same behavior. Hence, we use the combination $a_1=0.33$ and $t_0=0.14$ for all
  presented simulations in this work. However, it should be noted that our method converges for all possible choices of the free parameters.
 In Fig.\ \ref{convergence}, we demonstrate the convergence of the energy with respect to the number of propagators for the same system as in Fig.\ \ref{parameter}.
 However, since $V$ and $K$ nearly cancel for this particular combination of $r_s$, $\theta$ and $N$, we investigate the convergence of
 both contributions separately.
 The top panel shows the potential energy versus the inverse number of propagators $P^{-1}\propto\epsilon$ and we compare the PB-PIMC results 
 to the exact value (with the corresponding confidence interval) from CPIMC. We find that as few as two propagators allow for a
 relative accuracy $\Delta V/|V|\sim 10^{-4}$ and with $P=4$ the potential energy is converged within error bars.
In the bottom panel, we show the same information for the kinetic energy $K$. The variance of $K$ is one order of magnitude larger than that of $V$ and, for two propagators, we find the relative time step error $\Delta K/K\sim 10^{-3}$. With increasing $P$, the PB-PIMC results are fluctuating around the exact value, within error bars.

     \begin{figure}
      \centering
      \includegraphics[width=0.98\textwidth]{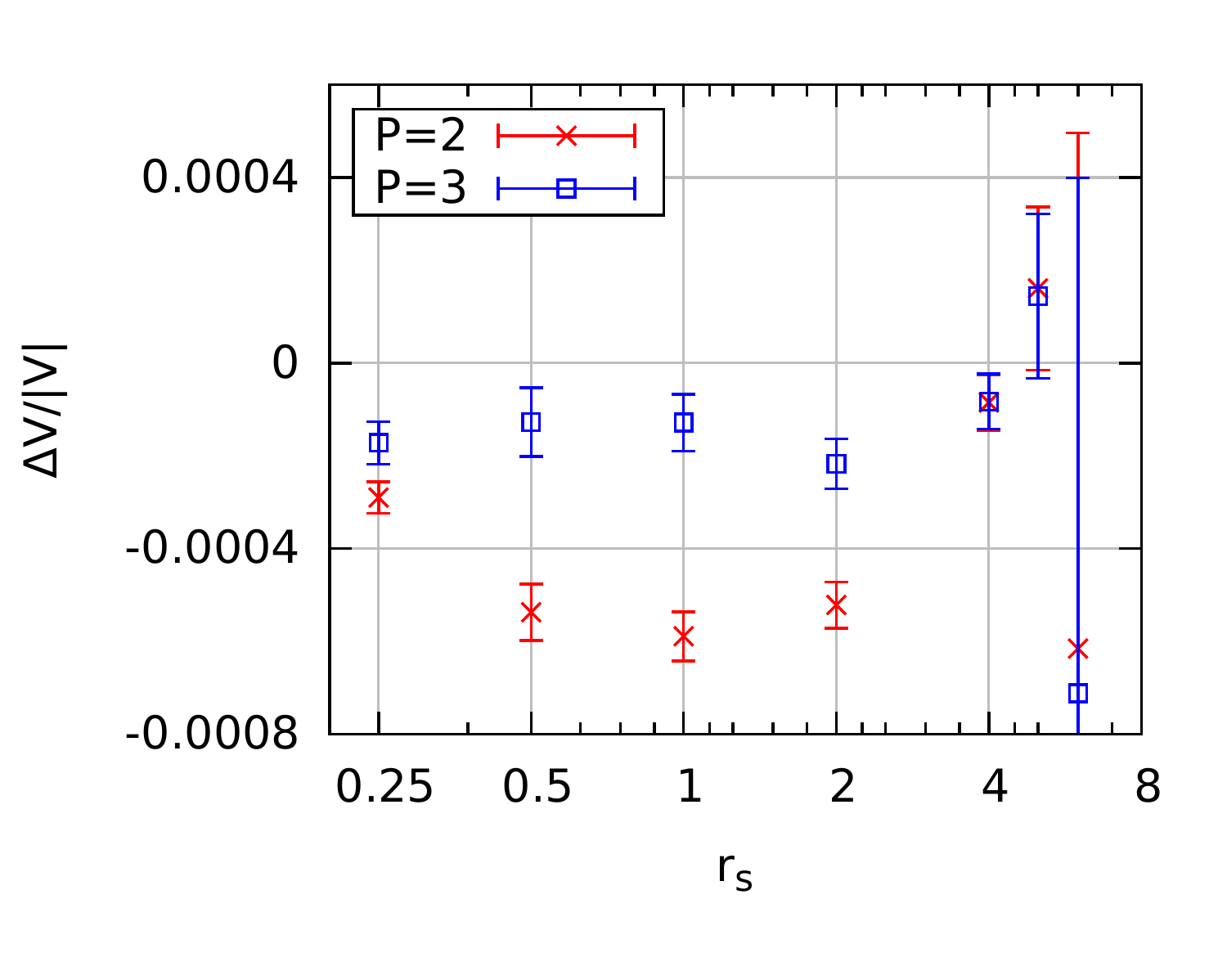}
      \caption{Accuracy of two and three propagators over a broad $r_s$ range for $N=4$ and $\theta=0.5$ with $t_0=0.14$ and $a_1=0.33$.
      We show the relative difference between the potential energy from PB-PIMC and CPIMC, $\Delta V/|V|$, for the optimal parameters from the fourth order
      propagator.
      }
      \label{pdiff}\vspace*{-0.2cm}
  \end{figure}
 
 Finally, we adress the $r_s$--dependence of the time step error by comparing PB-PIMC results for $V$ with $P=2$ (red crosses) and $P=3$ (blue squares)
 to the exact values from CPIMC.
 In Fig.\ \ref{pdiff}, the relative error of the potential energy $\Delta V/|V|$ is plotted versus $r_s$ for $N=4$ spin-polarized electrons at $\theta=0.5$.
 The increased errorbars for larger $r_s$ are a manifestation of the sign problem from CPIMC \cite{tim2}, while for the rest the statistical
 uncertainty from PB-PIMC predominates.
The time step error is smaller for three propagators over the entire $r_s$--range, as it is expected, and adopts a maximum
around $r_s=1$. This can be understood by recalling the source of the systematic error in PB-PIMC.
For $r_s\to0$, the UEG approaches an ideal system and the commutator error from $\hat{K}$ and $\hat{V}$ vanishes.
For $r_s\to\infty$, on the other hand, the particles are more separated and the system becomes more classical.
Therefore, the neglected commutator terms are most important at indermediate $r_s$, which is the case for the results in Fig.\ \ref{pdiff}.

We conclude that as few as two or three propagators provide sufficient accuracy to assess the discrepancy between CPIMC and RPIMC observed in previous studies \cite{prl}.
In particular, the selected benchmark temperature, $\theta = 0.5$, is even lower than for all other simulations to be presented in this work.
Hence, the observed time step error constitutes an upper bound for the accuracy of our results in the remainder of the paper.

 \subsection{Density parameter dependence\label{sdensity}}
  \begin{figure}
      \centering
      \includegraphics[width=0.98\textwidth]{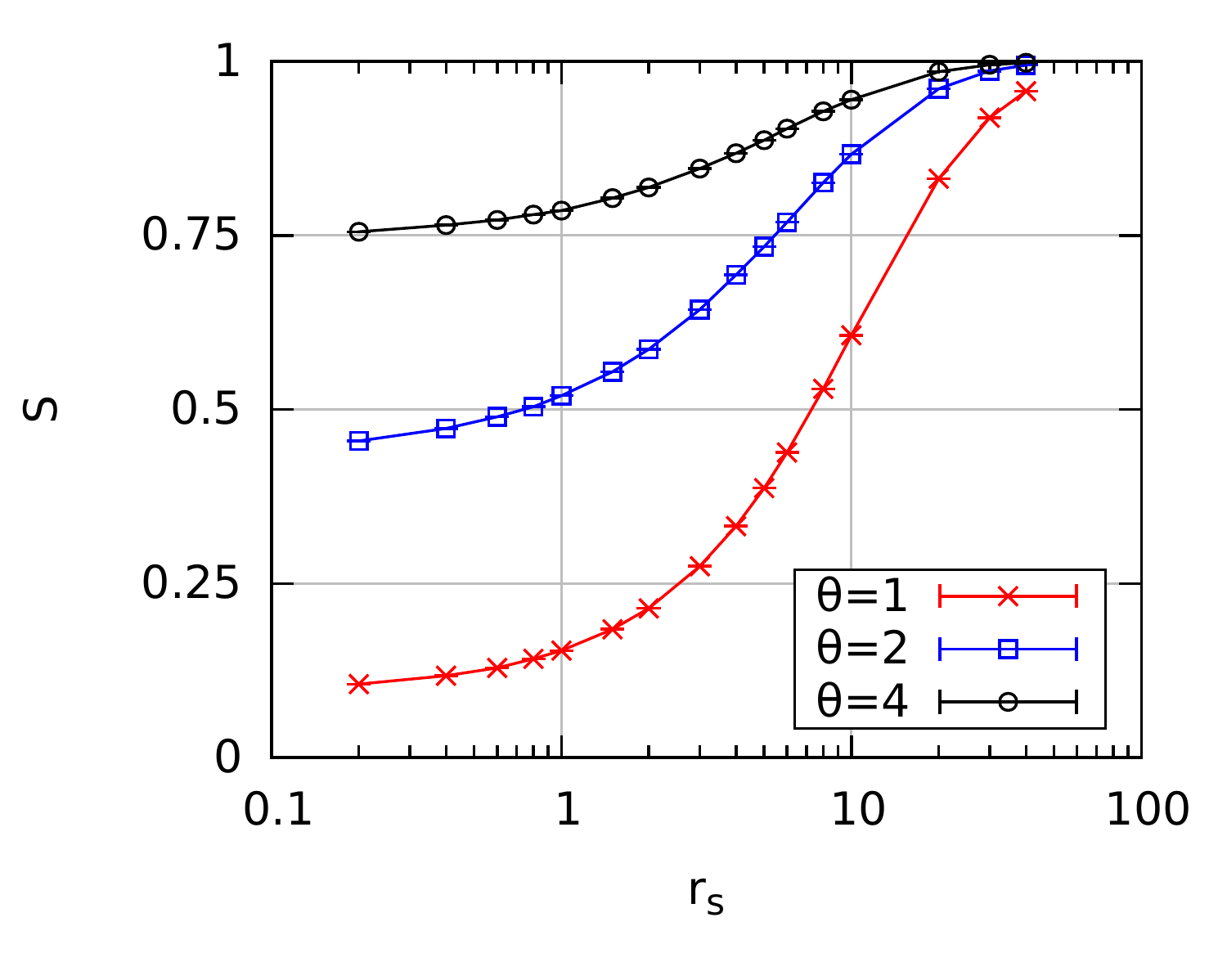}
      \caption{The average sign is plotted versus the density parameter $r_s$ for three different temperatures and $N=33$ spin-polarized electrons with $P=2$, $a_1=0.33$ and $t_0=0.14$.
      }
      \label{s1}\vspace*{-0.5cm}
  \end{figure}

    \begin{figure*}
      \centering
      \includegraphics[width=0.34\textwidth]{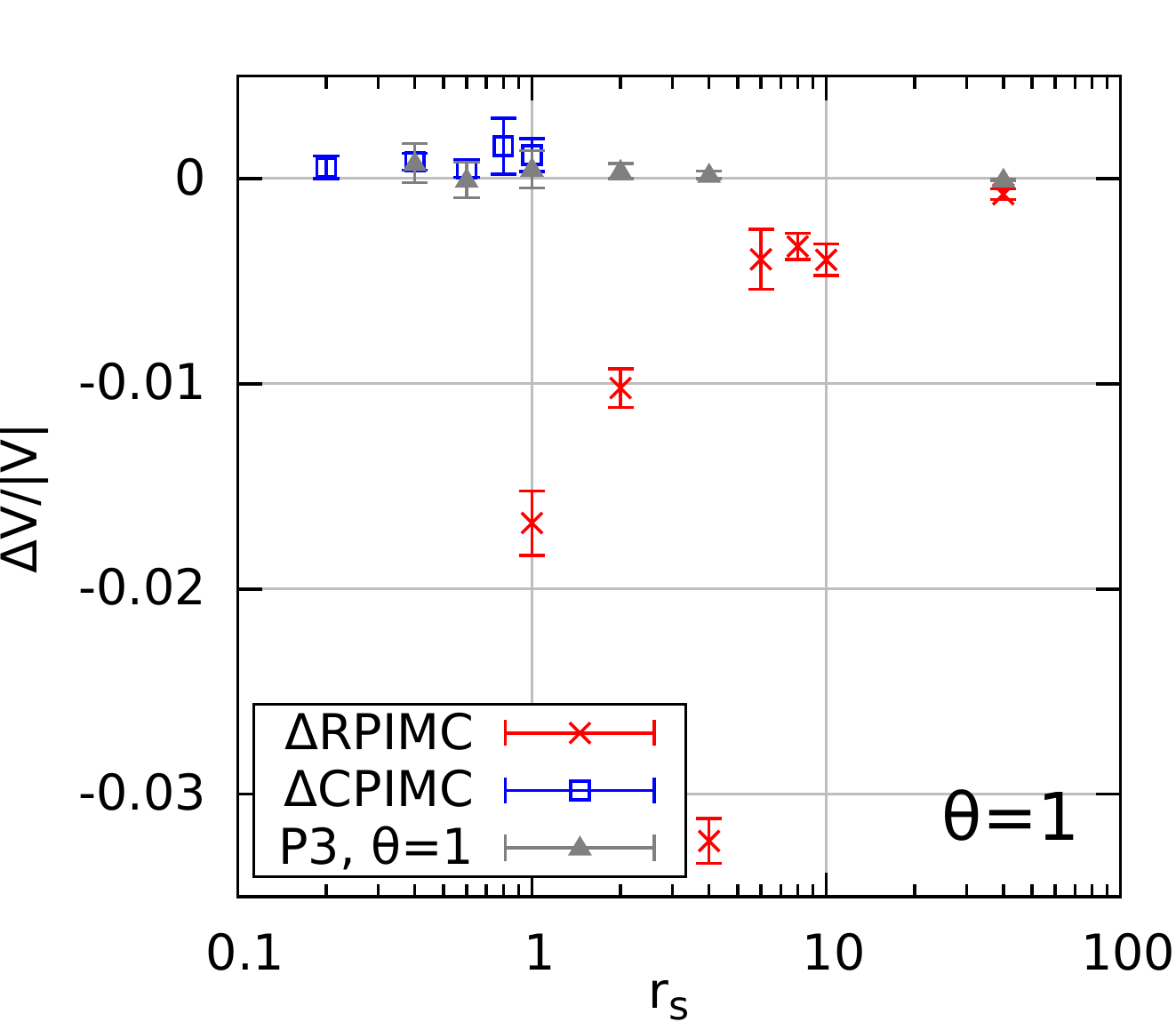}\hspace{-0.4cm}
      \includegraphics[width=0.34\textwidth]{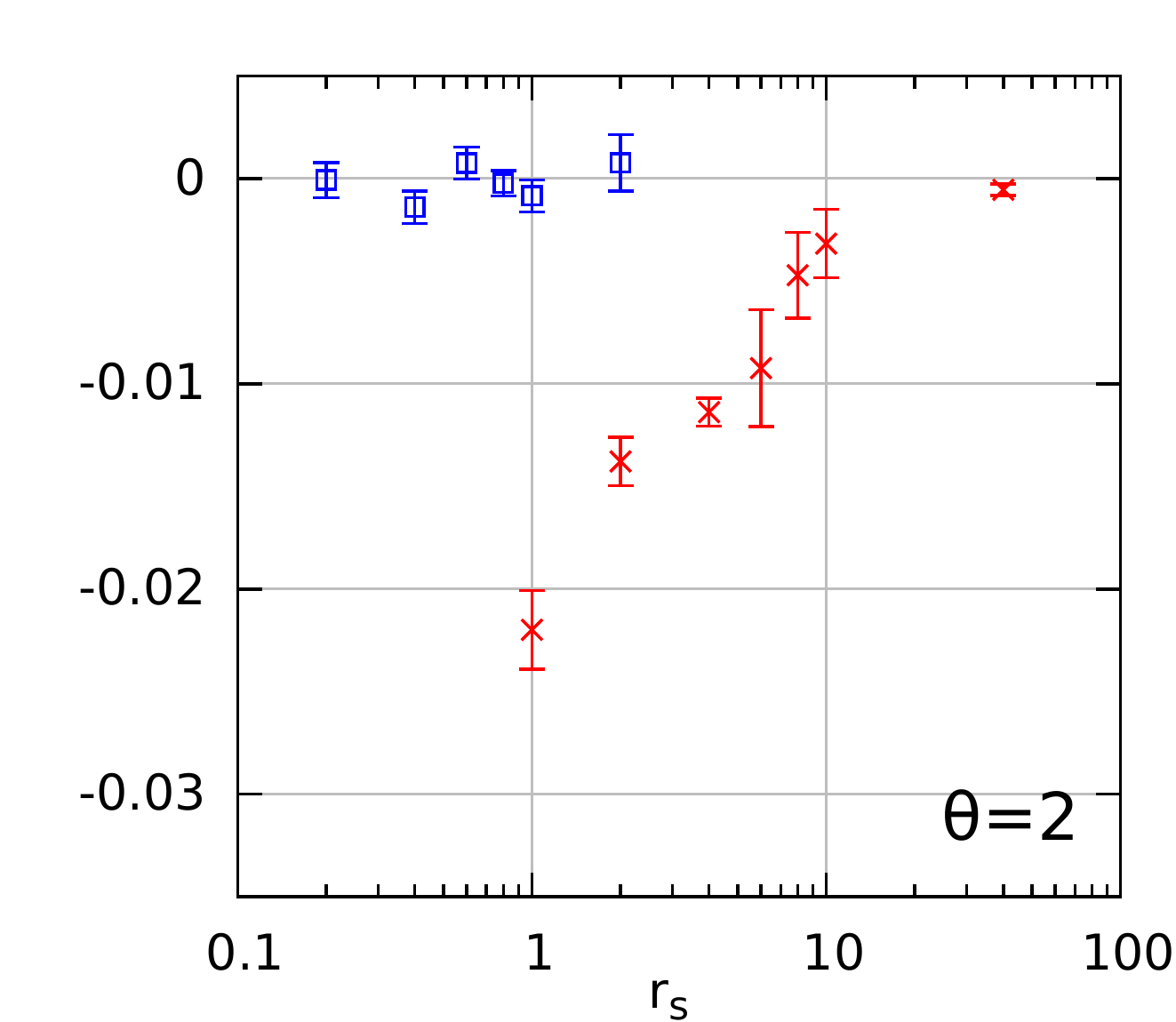}\hspace{-0.4cm}
      \includegraphics[width=0.34\textwidth]{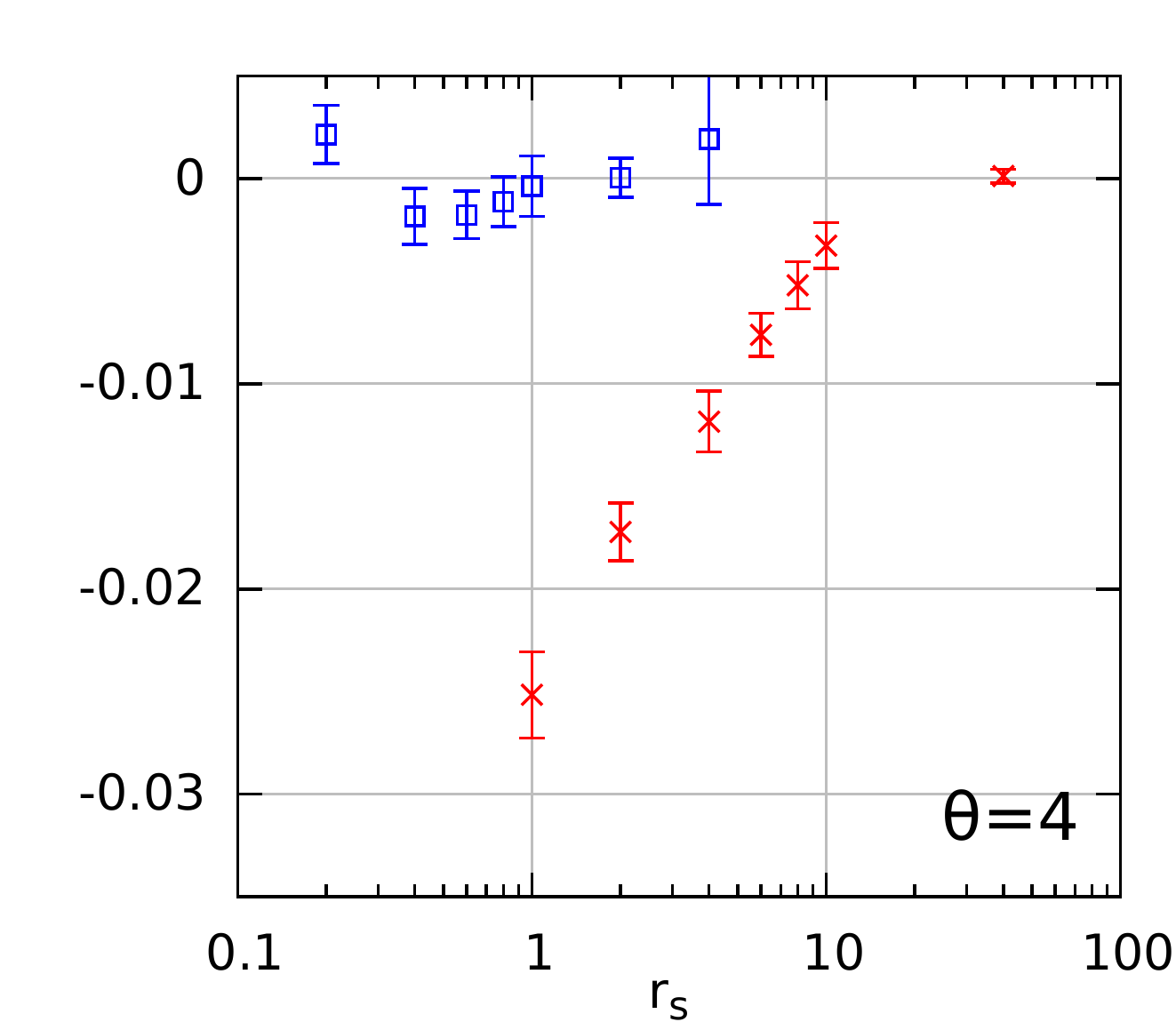}
      \includegraphics[width=0.34\textwidth]{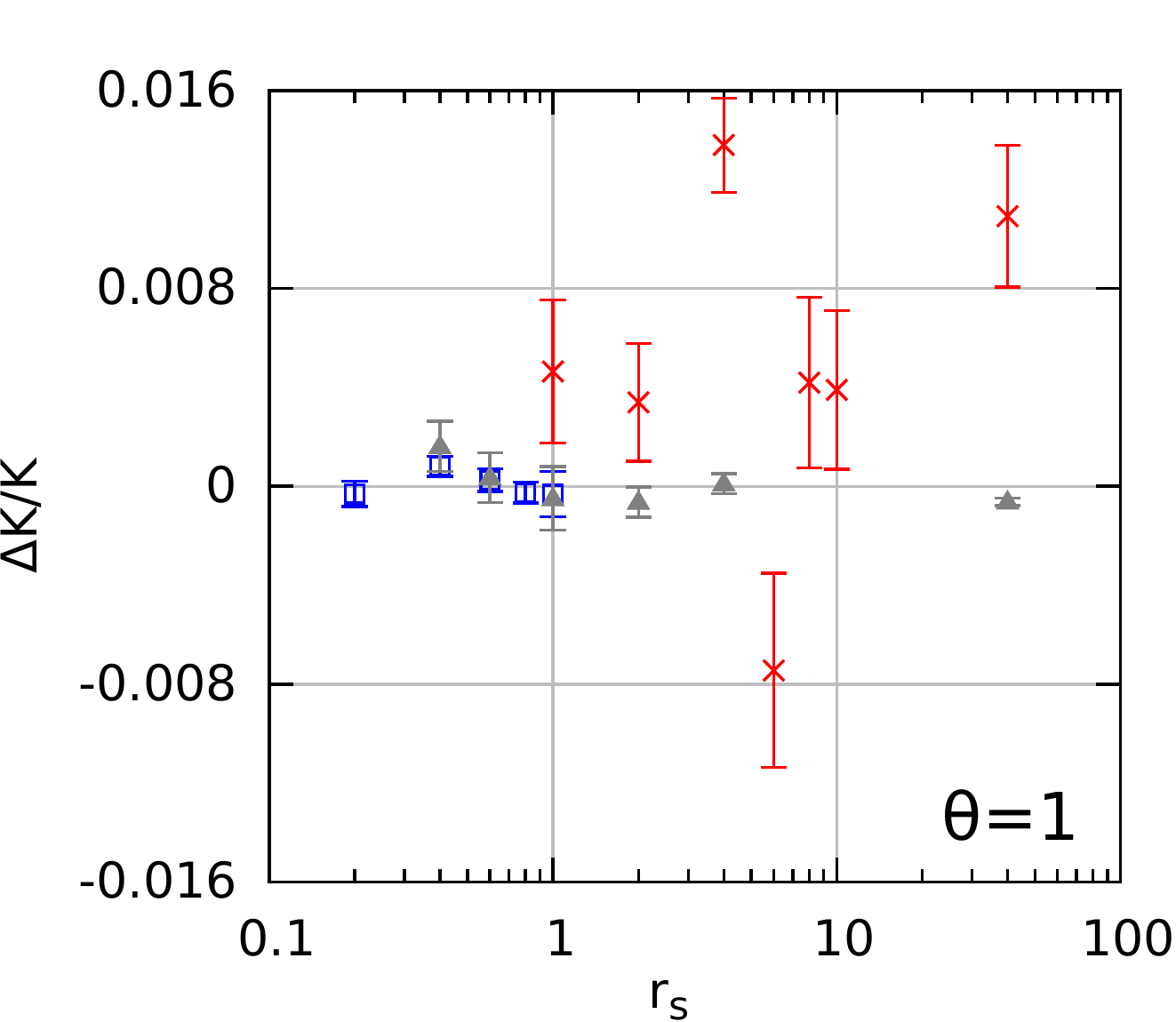}\hspace{-0.4cm}
      \includegraphics[width=0.34\textwidth]{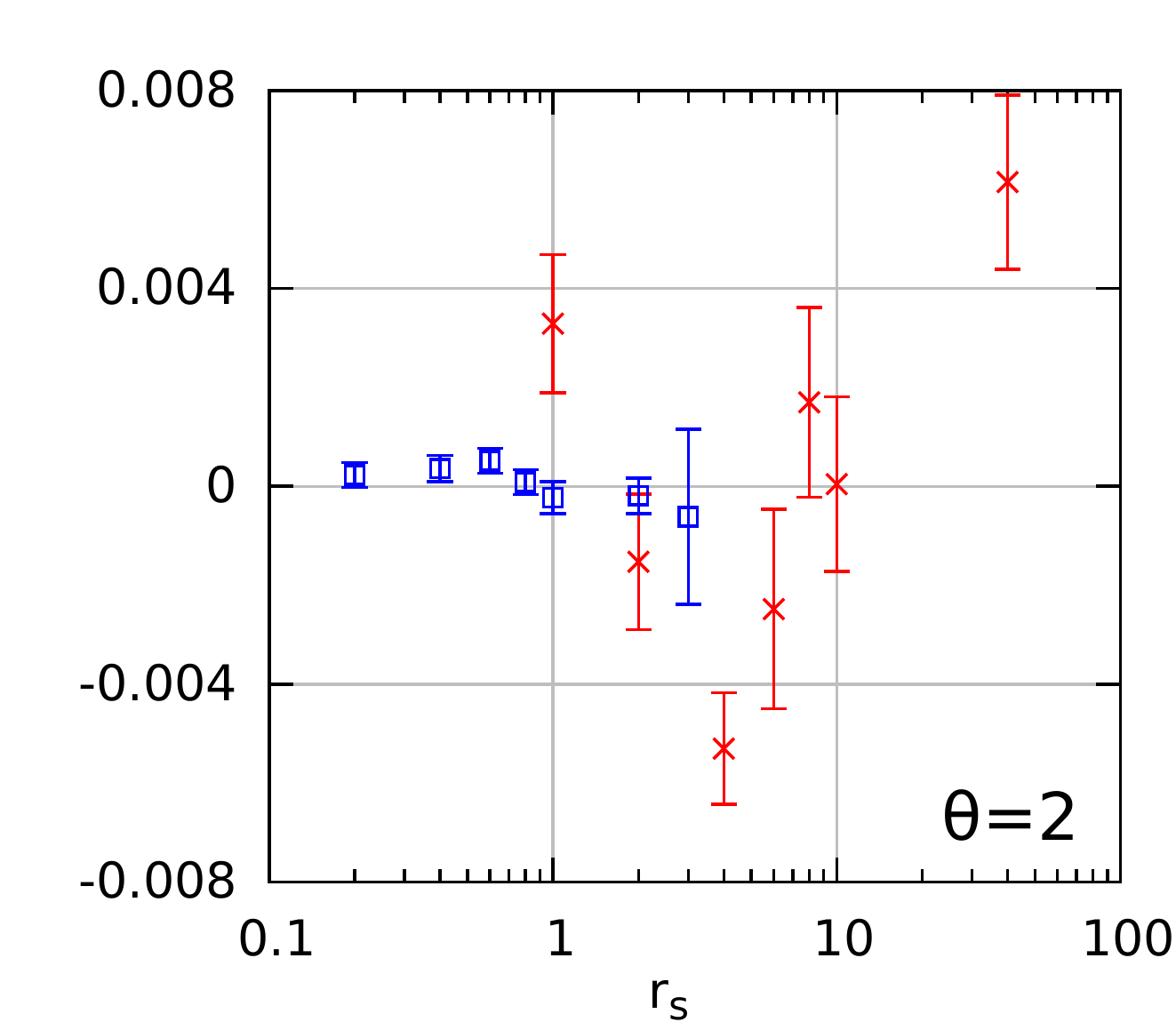}\hspace{-0.4cm}
      \includegraphics[width=0.34\textwidth]{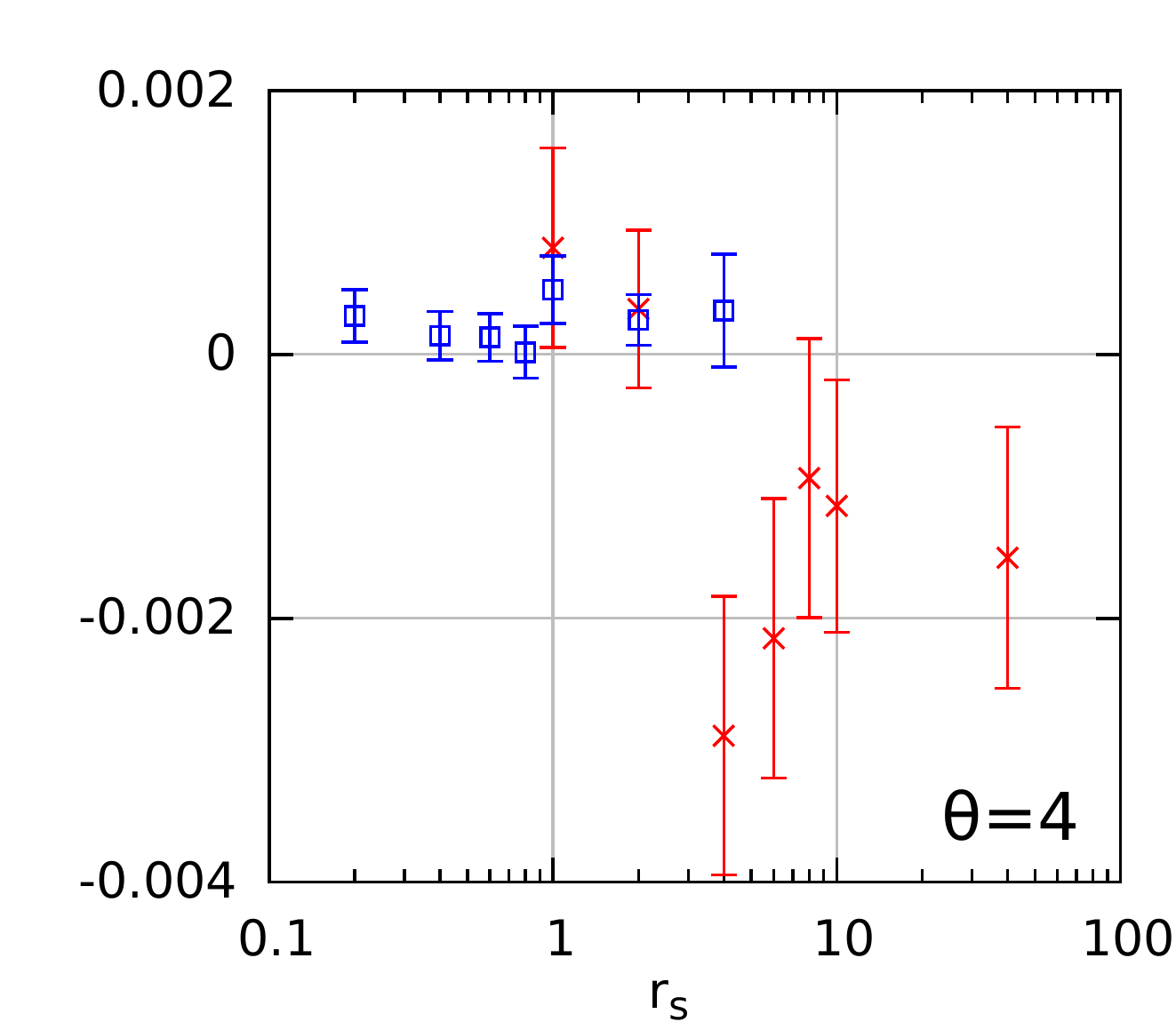}
      \includegraphics[width=0.34\textwidth]{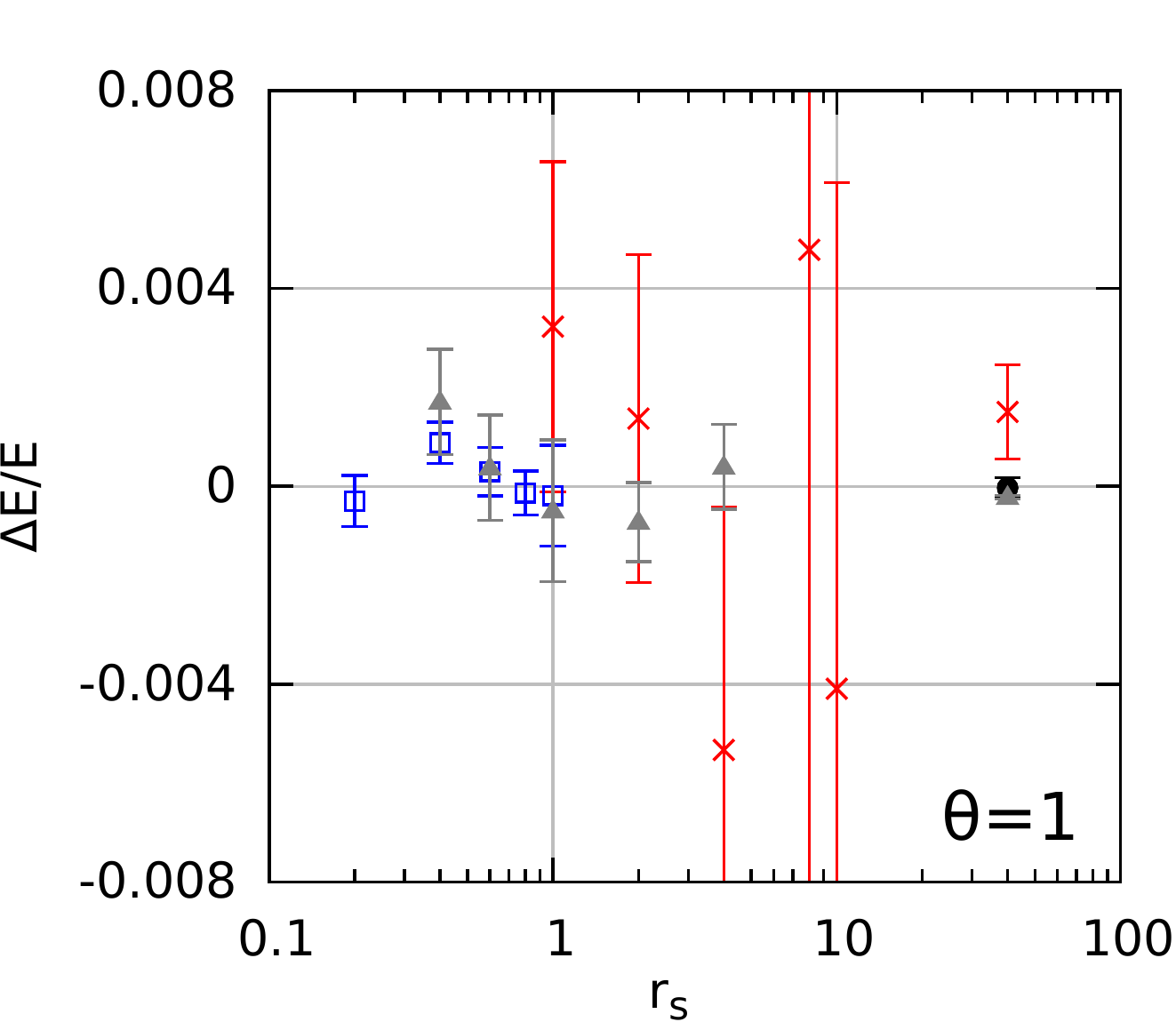}\hspace{-0.4cm}
      \includegraphics[width=0.34\textwidth]{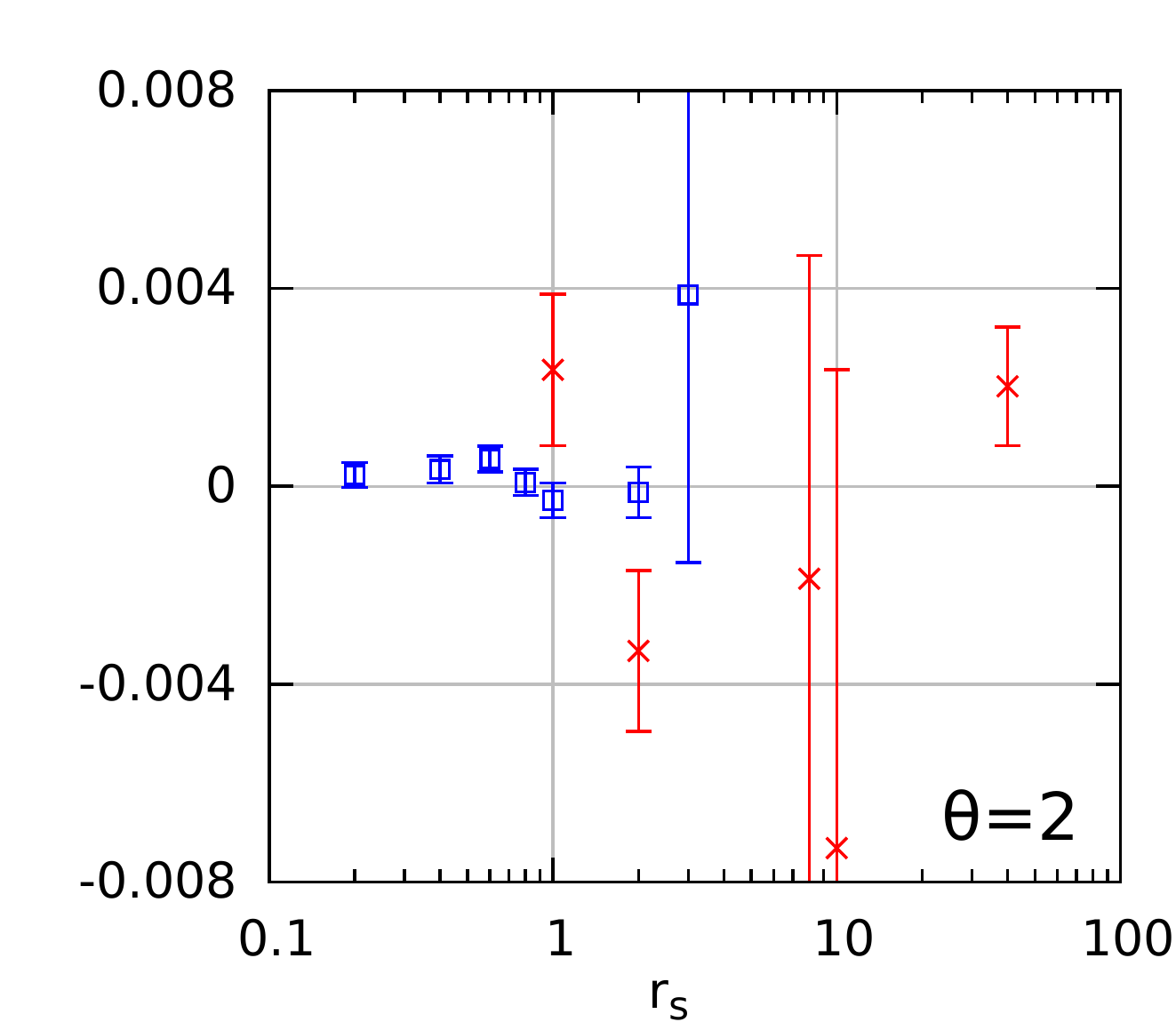}\hspace{-0.4cm}
      \includegraphics[width=0.34\textwidth]{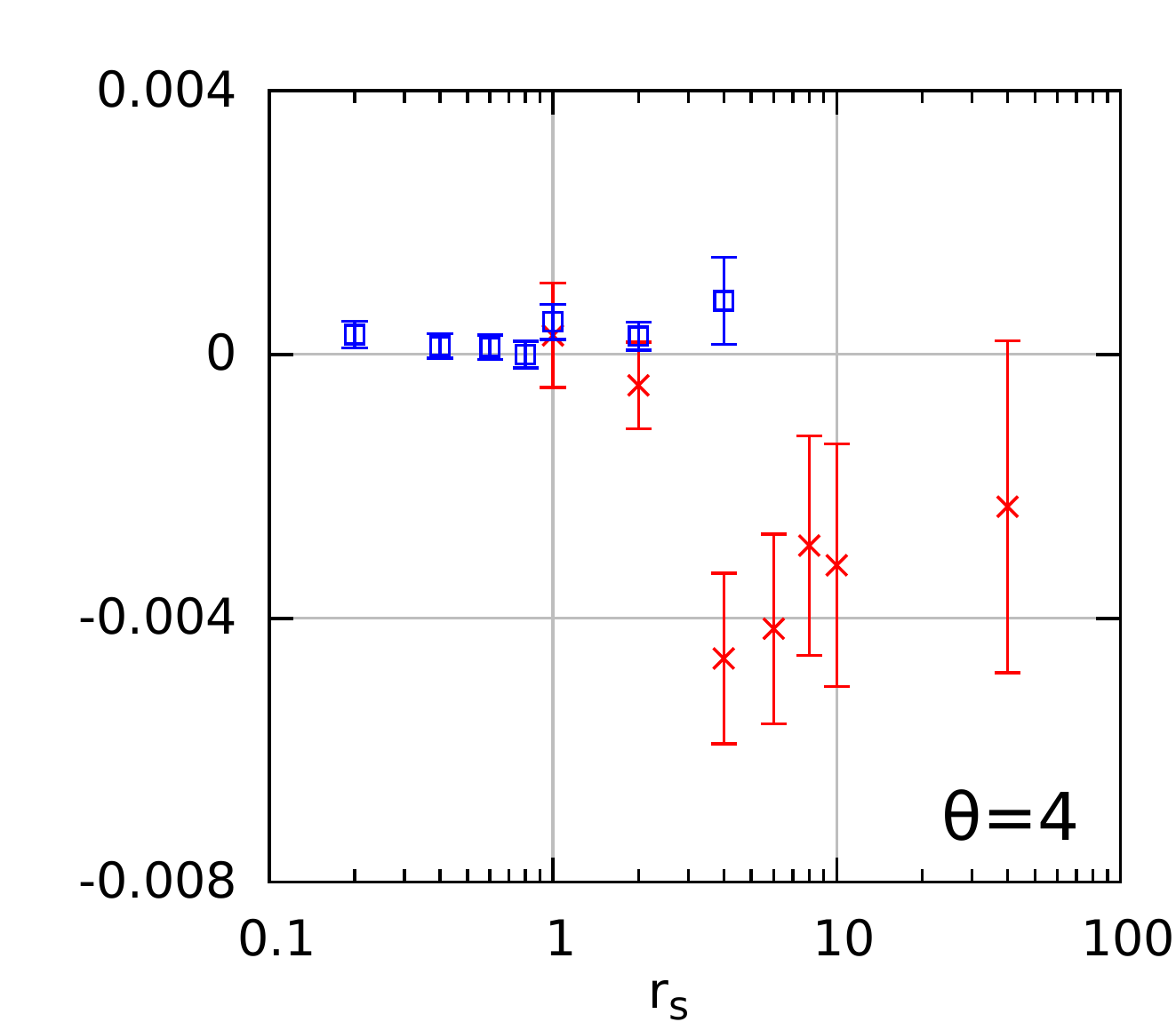}
      \caption{Comparison of PB-PIMC with CPIMC and RPIMC for $N=33$ spin-polarized electrons and three temperatures.
      In the top row, the relative deviation of the potential energy towards PB-PIMC with $P=2$, $t_0=0.14$ and $a_1=0.33$ is plotted
      versus $r_s$. The center and bottom rows display the same information for the kinetic and total energy, respectively.
      }
      \label{d1}
  \end{figure*}
 Among the most interesting questions regarding the implementation of PB-PIMC for the UEG is the range of applicability with respect to the density
 parameter $r_s$. To adress this issue, we simulate $N=33$ spin-polarized electrons, which corresponds to a closed momentum shell and is often used
 as a starting point for finite size corrections.
 In Fig.\ \ref{s1}, we show the average sign $S$ versus $r_s$ for three different temperatures over a broad density range.
 All curves exhibit a qualitatively similar behavior, that is, a smooth decrease of $S$ towards smaller $r_s$ until it saturates.
 At large $r_s$, the coupling induced particle separation mostly exceeds the extension of the single particle wavefunctions and quantum exchange effects do not play a dominant role.
 With decreasing $r_s$, the UEG approaches an ideal system and the particles begin to overlap, which leads to sign changes in the determinants.
 However, due the blocking, the average sign, instead of dropping exponentially, remains finite which implies that, for the three depicted temperatures, PB-PIMC is applicable over the entire density range. This is in stark 
 contrast to standard PIMC, see e.g.\ supplement of \cite{brown}.
 Nervertheless, with decreasing temperature the sign drops and the FSP makes the simulations more involved, cf.\ section \ref{td}.\par
 In Fig.\ \ref{d1}, we compare the corresponding energies with RPIMC \cite{brown2} and CPIMC \cite{prl}, where they are available.
The top row displays the relative difference in the potential energy towards PB-PIMC with two propagators.
For $\theta=4$ and $\theta=2$, we find excellent agreement with CPIMC. For the lowest temperature, $\theta=1$, 
the CPIMC values are systematically lower by $\Delta V/|V| \lesssim 10^{-3}$.
However, this discrepancy can be explained by the convergence behavior of the propagator, cf.\ Fig.\ \ref{pdiff}, since the potential (and kinetic)
energy is expected to converge from above towards the exact result.
To confirm this assumption, we also plot results for $P=3$ and $\theta=1$, visualized by the grey triangles. 
Evidently, these points coincide with the CPIMC data everywhere within the errorbars and, thus, can be regarded as quasi-exact.
The RPIMC data for $V$, on the other hand, exhibit a systematic discrepancy with respect to PB-PIMC and CPIMC \cite{prl}.
At $r_s=1$, the energies approximately differ by $\Delta V/|V|\sim 0.02$, but the difference decreases with increasing $r_s$.
In the center row, we display the relative difference in the kinetic energy.
Again, all PB-PIMC results are in good agreement with CPIMC.
On the other hand, there is no clear systematic deviation between the PB-PIMC and RPIMC data, although most RPIMC-values for $\theta=1$ are lower while the opposite holds for most values for $\theta=4$.
Finally, the bottom row displays the relative difference in the total energy. 
Interestingly, for $\theta=1$ the difference of RPIMC in $V$ and $K$ towards PB-PIMC nearly cancels, so that $E$ appears to be in good agreement.
In particular, even the value for $\theta=1$ and $r_s=4$, where the potential energy is an outlier, and both $V$ and $K$ exhibit a maximum deviation,
is almost within single error bars.
For completeness, we have also included the total energy for $\theta=1$ and $r_s=40$ from standard PIMC \cite{brown2}, cf.\ the black cirlce, which 
is in excellent agreement with PB-PIMC as well.
For $\theta=2$ and $\theta=4$, most RPIMC values for $E$ are higher than PB-PIMC, although the deviation hardly exceeds twice the error bars.

%In summary, PB-PIMC and CPIMC deliver - independently - identical results for $r_s\lesssim2$, while RPIMC, in particular for the potential energy,
%exhibits a significant deviation from the other methods.

 \subsection{Temperature dependece\label{td}}

   \begin{figure}
      \centering
      \includegraphics[width=0.98\textwidth]{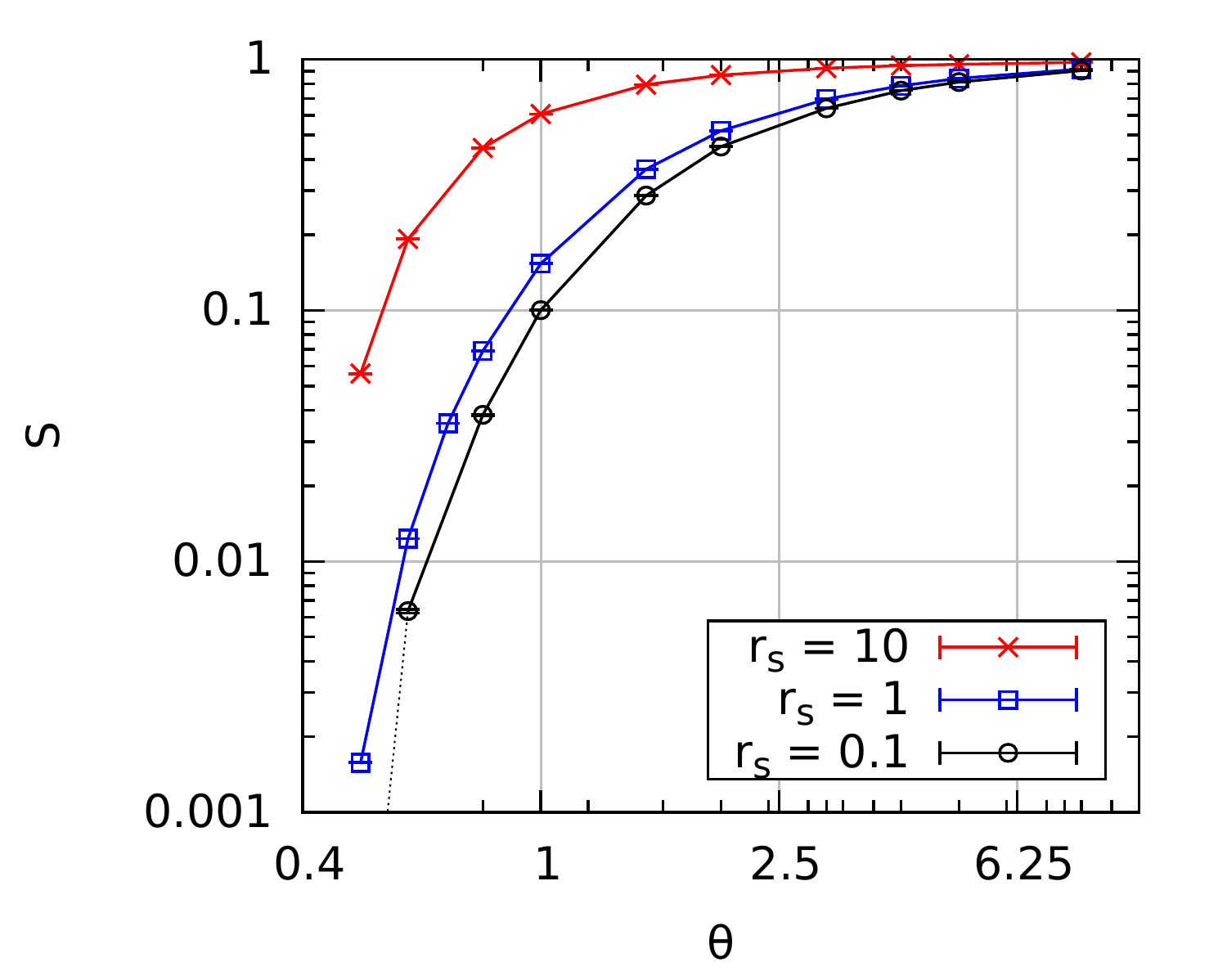}
      \caption{The average sign is plotted versus the temperature $\theta$ for $r_s=10$, $r_s=1$ and $r_s=0.1$ and $N=33$ spin-polarized electrons with $P=2$ and the free parameters $t_0=0.14$ and $a_1=0.33$.
      }
      \label{p1}\vspace*{-0.2cm}
  \end{figure}

 Finally, we investigate the performance of PB-PIMC with respect to the temperature.
 In Fig.\ \ref{p1}, the average sign is plotted versus $\theta$ for $N=33$ spin-polarized electrons at $r_s=10$, $r_s=1$ and $r_s=0.1$.
 All three curves exhibit a similar behavior, that is, a large sign $S$ at high temperature and a monotonous decay for
 $T\to0$. However, for $r_s=10$, the system is significantly less degenerate than for both other density parameters, and even 
 at $\theta=0.5$, the average sign of $S\approx0.056$ indicates that the simulations are feasible.
 For $r_s=1$ and $r_s=0.1$, the decay of $S$ is more rapid and, at low temperature, the simulations are more involved.
 In particular, half the Fermi temperature seems to constitute the current limit down to which reasonable results can be achieved
 for such $r_s$--values (and this particle number) and, for $r_s=0.1$, the sign is zero within error bars, cf.\ the dashed line.
 Finally, we note that the average signs for the two smaller depicted $r_s$ parameters are more similar to each other than to $r_s=10$.
 We characterize the temperature in units of the ideal Fermi temperature, which is appropriate for weak coupling. However, for large $r_s$, the system becomes increasingly
 nonideal and, therefore, $\theta$ does not constitute an adequate measure for the degeneracy.

   \begin{figure}
      \centering
      \includegraphics[width=0.98\textwidth]{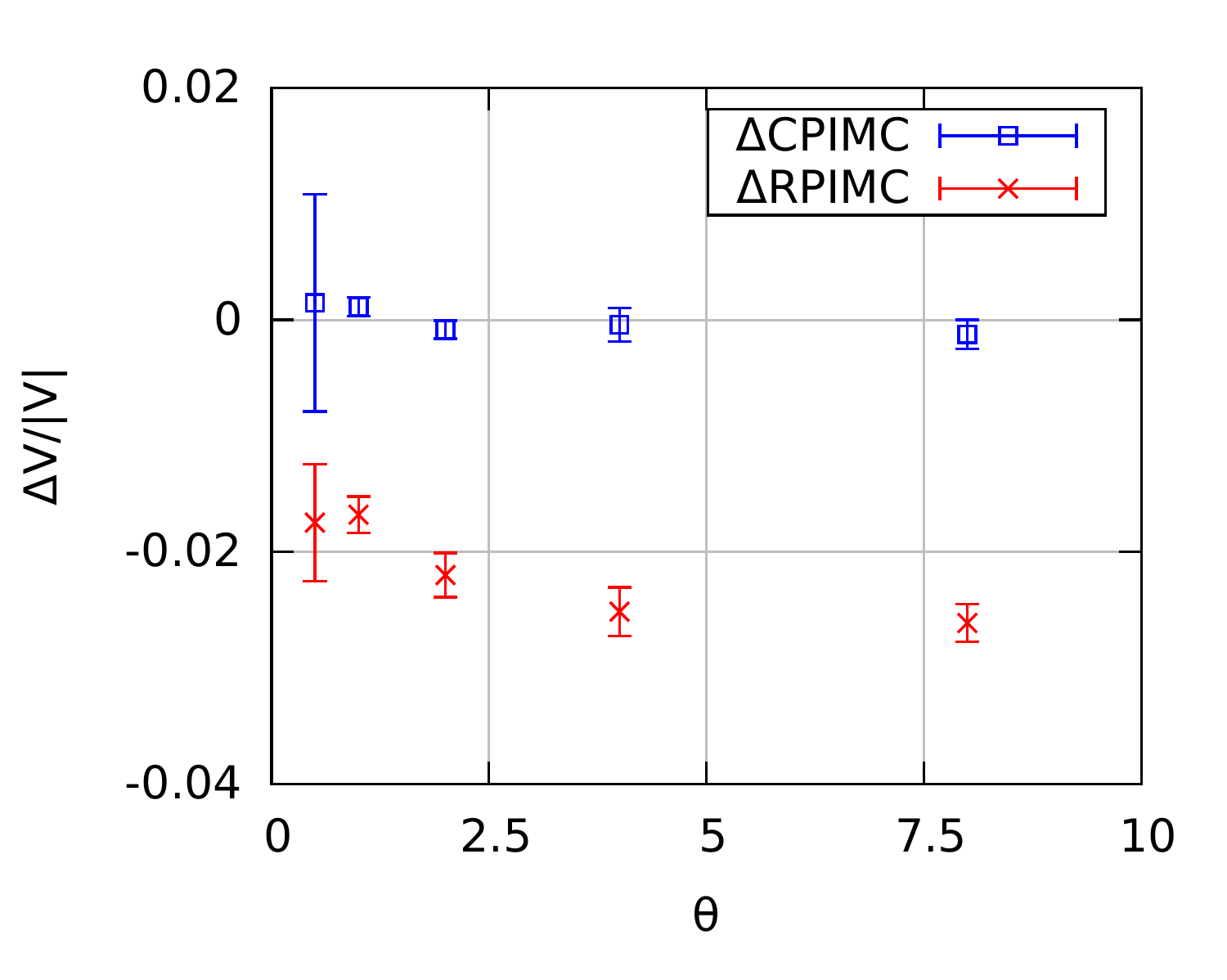}\\
       \vspace*{-0.9cm}
      \includegraphics[width=0.98\textwidth]{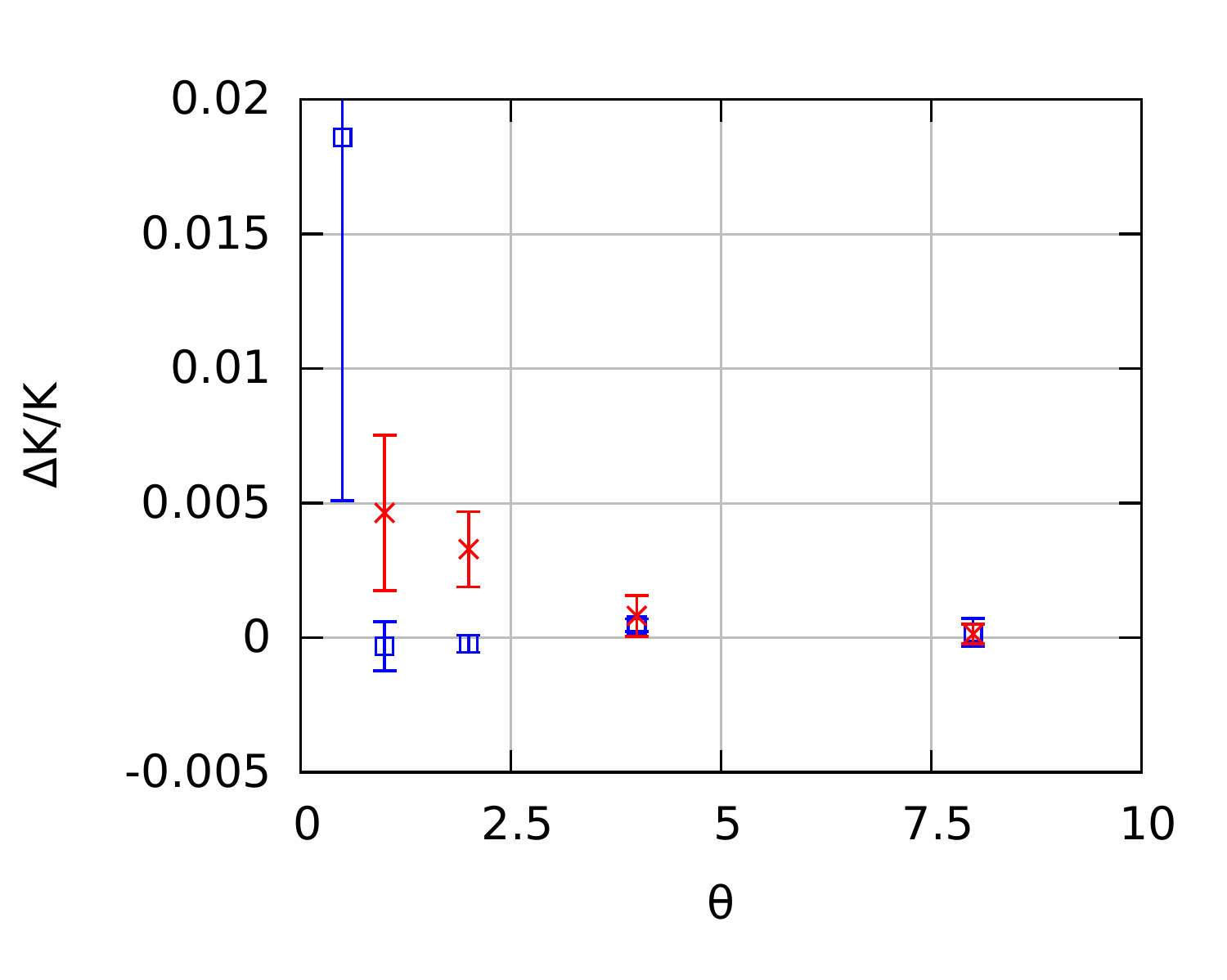}
      \caption{Comparison with CPIMC and RPIMC as a function of temperature. In the top panel, the relative deviation of the potential energy
      from the PB-PIMC result is plotted versus $\theta$ for $N=33$ spin-polarized electrons and $r_s=1$. The bottom panel displays the same information
      for the kinetic contribution.
      }
      \label{c1}\vspace*{-0.5cm}
  \end{figure}

 In Fig.\ \ref{c1}, we compare the energies of the $N=33$ electrons at $r_s=1$ from PB-PIMC both to RPIMC \cite{brown2} and CPIMC.
 The top panel displays the relative difference in the potential energy versus $\theta$.
 The CPIMC results for $V$ are in good agreement with PB-PIMC, while the RPIMC data are systematically higher, by about $2\%$.
 Interestingly, this behavior appears to be almost independent of the temperature.
 In the bottom panel, the same information is shown for the kinetic energy and, again, PB-PIMC agrees with CPIMC over the entire temperature range.
 The large statistical uncertainty at $\theta=0.5$ is a manifestation of the FSP in PB-PIMC, which prevents us from obtaining more precise kinetic
 energies with feasible computational effort. The RPIMC data for $K$ are
 slightly lower, at low temperature, which confirms the trend observed by Schoof \textit{et al.} \cite{prl}, and seems to converge towards the other methods for large $\theta$.

\section{Discussion}
 In summary, we have successfully extended the permutation blocking path integral Monte Carlo (PB-PIMC) method \cite{dornheim} to the uniform electron gas at finite temperature.
 We have started the discussion with a brief introduction of our simulation scheme, which combines a fourth-order factorization of the density matrix
 with the application of antisymmetric imaginary time propagators, i.e., determinants.
 This allows us to combine permutations, which appear as individual configurations with positive and negative sign in standard PIMC,
 into a single configuration weight. Therefore, the average sign in our simulations is significantly increased. 
 
 To assert the quality of our numerical results, we have investigated the optimization of the free parameters of our propagator
 and demonstrated the convergence of both the potential and kinetic energy with respect to the number of imaginary time steps.
 We have found that even for the lowest considered temperature, $\theta=0.5$, as few as two propagators allow for a relative
 accuracy of $0.1\%$ and $0.01\%$ in the kinetic and potential energy, respectively. After this preparatory work,
 we have shown results for $N=33$ spin-polarized electrons, which is a commonly used model system as it is well suited to be a starting point for 
 the extrapolation to the macroscopic limit (finite size corrections).
 Interestingly, PB-PIMC is feasible over the entire density range and, therefore, allows us to compare our results to both CPIMC and RPIMC data, where they are available. 
% In particular, we are able to resolve a previously reported discrepancy between the two approaches around $r_s=1$.
 Our PB-PIMC data exhibit a very good agreement with CPIMC, for both the potential and kinetic energy,
 for all three investigated temperatures. 
 On the other hand, we observe deviations between PB-PIMC and RPIMC of up to $3\%$ in the potential energy, which
 decreases towards strong coupling. For the kinetic energy, we find no systematic trend although, for $\theta=1$, most of the RPIMC-values are smaller while, for $\theta=4$, most are larger than the PB-PIMC results. However, for both temperatures this deviation hardly exceeds twice the RPIMC errorbars.
 
 Finally, we have investigated the applicability of PB-PIMC to the $N=33$ spin-polarized electrons with respect to the temperature.
 With decreasing $\theta$, exchange effects lead to more negative determinants in the configuration weights and, therefore, 
 a smaller average sign. For the physically most interesting density regime, $r_s\sim 1$, simulations are feasible above $\theta=0.5$ while
 for larger $r_s$ even lower temperatures are possible.
 A comparison of the energies for $r_s=1$ over the entire applicable temperature range has again revealed an excellent agreement with CPIMC. On the other hand, we observe a nearly $\theta$-independent relative deviation between PB-PIMC and RPIMC in the potential energy of approximately $2\%$, whereas differences in the kinetic energy are observed only towards low temperature.

 We conclude that our permutation blocking PIMC approach is capable to provide accurate results for the UEG over a broad parameter range. This approach is efficient above a minimum temperature of about $0.5 T_F$ and, thus, complements CPIMC.
 Even though PB-PIMC carries a small systematic error (which is controllable and depends only on the number of time slices),  we expect it to be useful for the development and test of other new techniques such as DMQMC \cite{blunt,malone} and other novel versions of fermionic PIMC, such as the approximate treatment of exchange cycles by DuBois \textit{et al.}\ \cite{dubois} or a variational approach to the RPIMC nodes, e.g.\ \cite{brownphd}. 
 
 A natural follow-up of this work will be the extension of PB-PIMC to unpolarized systems which, in 
 combination with CPIMC, should allow for a nearly complete description of the finite temperature UEG over the entire density range.
 In addition, we aim for the application or derivation of finite size corrections in order to extrapolate our results to the macroscopic limit \cite{fraser,drummond,lin}
 which could be followed by the construction of a new analytical fit formula for the UEG at finite temperature, e.g.\ \cite{brown3,karasiev}.
 Finally, since PB-PIMC allows for efficient simulations in the warm dense matter regime, applications to two-component
 plasmas, such as dense hydrogen \cite{bonitz,morales,proton}, are within reach.

\section*{Acknowledgements}
We acknowledge stimulating discussions with C. Hann (Kiel and Durham, North Carolina) and V.S. Filinov (Moscow). This work is supported by the Deutsche Forschungsgemeinschaft via SFB TR-24
project A9 and via project BO 1366/10 as well as by grant
SHP006 for CPU time at the Norddeutscher Verbund f\"ur
Hoch- und H\"ochstleistungsrechnen (HLRN).


\begin{thebibliography}{10}
\bibitem{fci} For completeness, we mention that improved ground state methods have been developed, the most accurate apparently being the full configuration interaction QMC method \cite{fci1,fci2,fci3}
\end{thebibliography}
\end{document}